\documentclass{mem}
\usepackage{natbib}\usepackage{txfonts}\usepackage{balance}
\usepackage{graphicx}
\usepackage[a4paper,breaklinks,dvipdfm]{hyperref}
\idline{1}{1}
\begin{document}
\def\teff{$T\rm_{eff }$}
\def\kms{$\mathrm {km s}^{-1}$}

\title{
The flattening of globular clusters: internal rotation or velocity anisotropy?
}

   \subtitle{}

\author{
P.\,Bianchini\inst{1,2} 
\and A. L.\,Varri\inst{1,3}
\and G.\,Bertin\inst{1}
\and A.\,Zocchi\inst{1}
          }

\institute{
Universit\`a degli Studi di Milano,
Dipartimento di Fisica, via Celoria 16, 20133 Milano, Italy
\and
Max-Planck Institute for Astronomy, K\"onigstuhl 17, 69117 Heidelberg, Germany
\and Indiana University, Department of Astronomy, 727 East 3rd Street, Swain West 319, Bloomington, IN 47405-7105, USA\\
\email{bianchini@mpia.de}
}

\authorrunning{Bianchini}

\titlerunning{Flattening due to internal rotation or velocity anisotropy?}

\abstract{Internal rotation is considered to play a major role in determining the structure and dynamics of some globular clusters. We present a dynamical analysis of the photometry and three-dimensional kinematics of 47 Tuc and $\omega$ Cen, by means of a new family of self-consistent axisymmetric rotating models. The combined use of line-of-sight velocities and proper motions allows us to obtain a global description of the internal dynamical structure of the objects together with an estimate of their dynamical distances.  
The well-relaxed cluster 47 Tuc is very well interpreted by our dynamical models; in particular, internal rotation is found to explain the observed morphology. For the partially relaxed cluster $\omega$~Cen, the selected model provides a good representation of its complex three-dimensional kinematics, in general qualitative agreement with the observed anisotropy profile, which is characterized by tangential anisotropy in the outer parts; discrepancies are found between the observed and the expected ellipticity profile and are ascribed to the presence of a high degree of radial anisotropy in the intermediate region and to its interplay with rotation.

}
\maketitle{}

\section{Introduction}
Globular clusters (GCs) have long been considered simple quasi-relaxed nonrotating stellar systems, characterized by spherical symmetry and isotropy in velocity space. Spherical nonrotating isotropic models (e.g., the \citealp{King1966} and the spherical \citealp{Wilson1975} models) have indeed provided an adequate zero-order description of their internal dynamics.

However, deviations from sphericity are observed and can now be measured in quantitative detail (see \citealp{Geyer1983}, \citealp{WhiteShawl1987}, and \citealp{ChenChen2010}). In addition, significant internal rotation has been detected in a growing number of GCs from line-of-sight velocity measurements \citep{Bellazzini2012} and, in a few cases, from proper motion measurements \citep[e.g.,][]{vanLeeuwen2000,Anderson2003}. Detailed three-dimensional kinematics is therefore available for selected GCs, in particular thanks to proper motion data sets from the Hubble Space Telescope \citep[e.g.,][]{Anderson2010}. This progress calls for the development of a more realistic dynamical modeling framework, in which rotation and deviations from sphericity are taken into consideration.

Interest in the study of internal rotation derives from a number of open problems: (1) internal rotation, together with external tides, and pressure anisotropy, is considered to be one of the main physical factors responsible for the observed flattening of GCs \citep{vandenBergh2008}; (2) the presence of global angular momentum is expected to change the long-term dynamical evolution of stellar systems \citep{Fiestas2006}; (3) differential rotation can cooperate with pressure anisotropy to produce nontrivial gradients in the kinematic profiles \citep{Varri2012}. Unfortunately, in only few cases has internal rotation been studied by a quantitative application of nonspherical rotating dynamical models.


Here we present, as a short preliminary report on an extensive study presented in a separate article (Bianchini et al. 2013, paper submitted), the application of a family of self-consistent rotating models recently constructed \citep{Varri2012} to two Galactic GCs, 47~Tuc and $\omega$~Cen. The dynamical models are compared to the photometric and three-dimensional kinematic data. Furthermore, by taking into consideration the inclination angle of the rotation axis of the stellar systems, we perform a detailed analysis of the morphology of the clusters, thus testing its connection with rotation. Finally, since the two clusters are in different relaxation states (47 Tuc with $\log T_c < 8$ falls in the class of well relaxed clusters, whereas $\omega$ Cen with $\log T_c > 9$ should be considered as only partially relaxed; here $T_c$ indicates the core relaxation time expressed in years, \citealp{Zocchi2012}) we can test whether internal rotation plays different roles in systems under different relaxation conditions.

\section{Dynamical models and fitting procedure}

The family of self-consistent axisymmetric models used in our analysis was specifically designed to describe quasi-relaxed stellar systems \citep{Varri2012}. These models are defined by a distribution function dependent on the integral of the motion {$I(E,J_z)=E-(\omega J_z)/(1+bJ_z^{2})$, 
such that the rotation is differential, that is, approximately rigid in the center and declining and eventually vanishing in the outer parts. The velocity dispersion tensor is characterized by isotropy in the central region, weak radial anisotropy in the intermediate regions, and tangential anisotropy in the outer parts (this behavior results from the requirement of self-consistency and from the adopted truncation prescription in phase space). Three dimensionless parameters define the models: the dimensionless depth of the potential well $\Psi$, the rotation-strength parameter, and the parameter $b$ defining the shape of the rotation profile. Furthermore, 5 additional quantities must be specified: 2 physical scales (e.g., the radial scale and the velocity scale), the mass-to-light ratio (to convert mass density profiles into surface brightness profiles), the inclination angle of the rotation axis with respect to the line of sight, and the distance to the cluster (to convert proper motions into km~s$^{-1}$). 
 
Models and observations are matched through a kinematic priority approach in the following steps. First, we determine the dimensionless parameters that characterize the internal structure of the models by following natural criteria suggested by the observed kinematics, in particular by the characteristics of differential rotation. Then, we set the physical scales by means of standard fits to the surface brightness profile and line-of-sight dispersion and rotation profiles. An additional fit to the proper motion dispersion profiles gives us an estimate for the distance of the cluster. Finally, once all the scales have been fixed, other observable quantities follow as predictions; in particular, the selected model makes a well defined prediction on the ellipticity profile.

\section{47 Tuc: a well-relaxed globular cluster}

\begin{figure}[b!]
\centering
\includegraphics[width=0.39\textwidth]{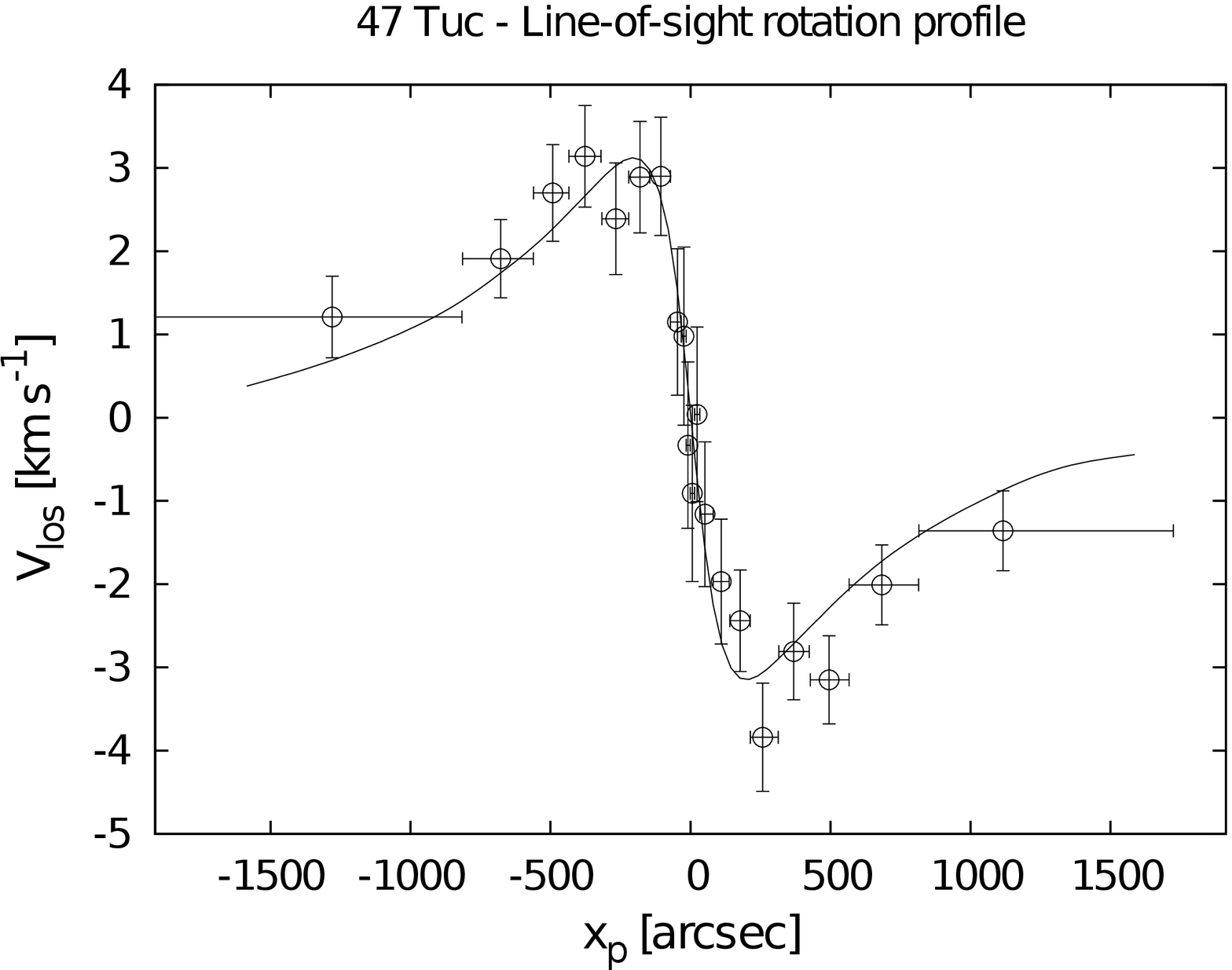}\\
\vspace{0.1cm}
\includegraphics[width=0.445\textwidth]{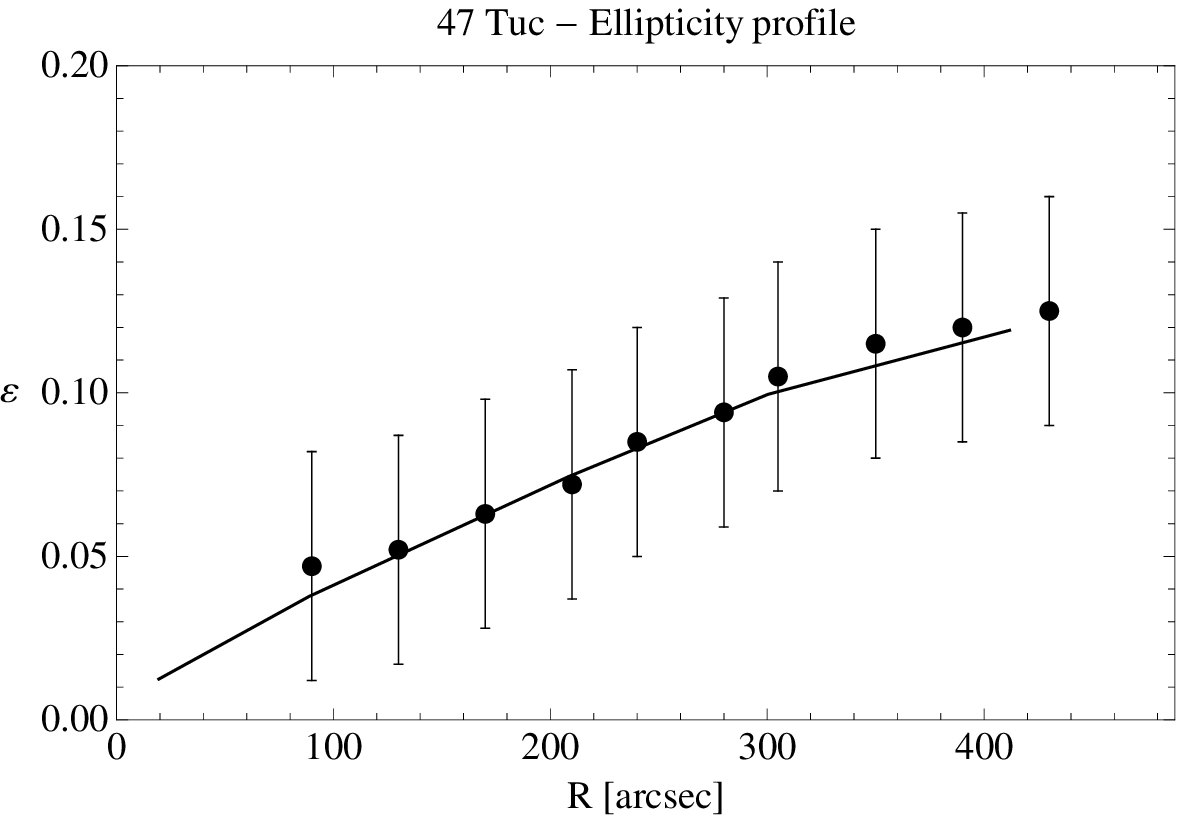}
\caption{\footnotesize Line-of-sight rotation profile measured along the major axis (top panel) and ellipticity profile (bottom panel) for 47 Tuc. Solid lines represent the model profiles, open circles the observational kinematic data points, and black dots the observed ellipticities from \citet{WhiteShawl1987}.
}
\label{47tuc}
\end{figure}
\begin{figure}[b!]
\centering
\includegraphics[width=0.39\textwidth]{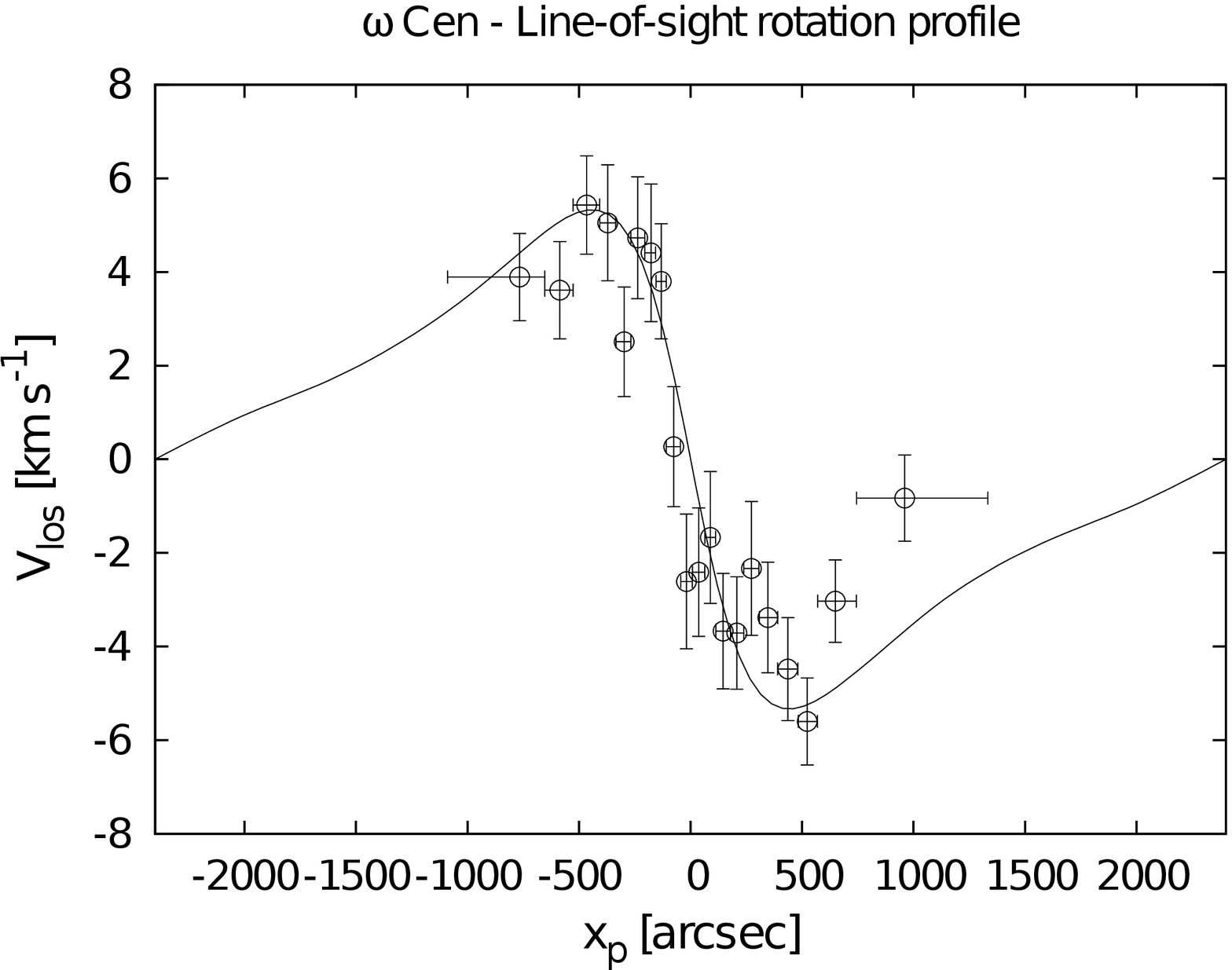}\\
\vspace{0.1cm}
\includegraphics[width=0.445\textwidth]{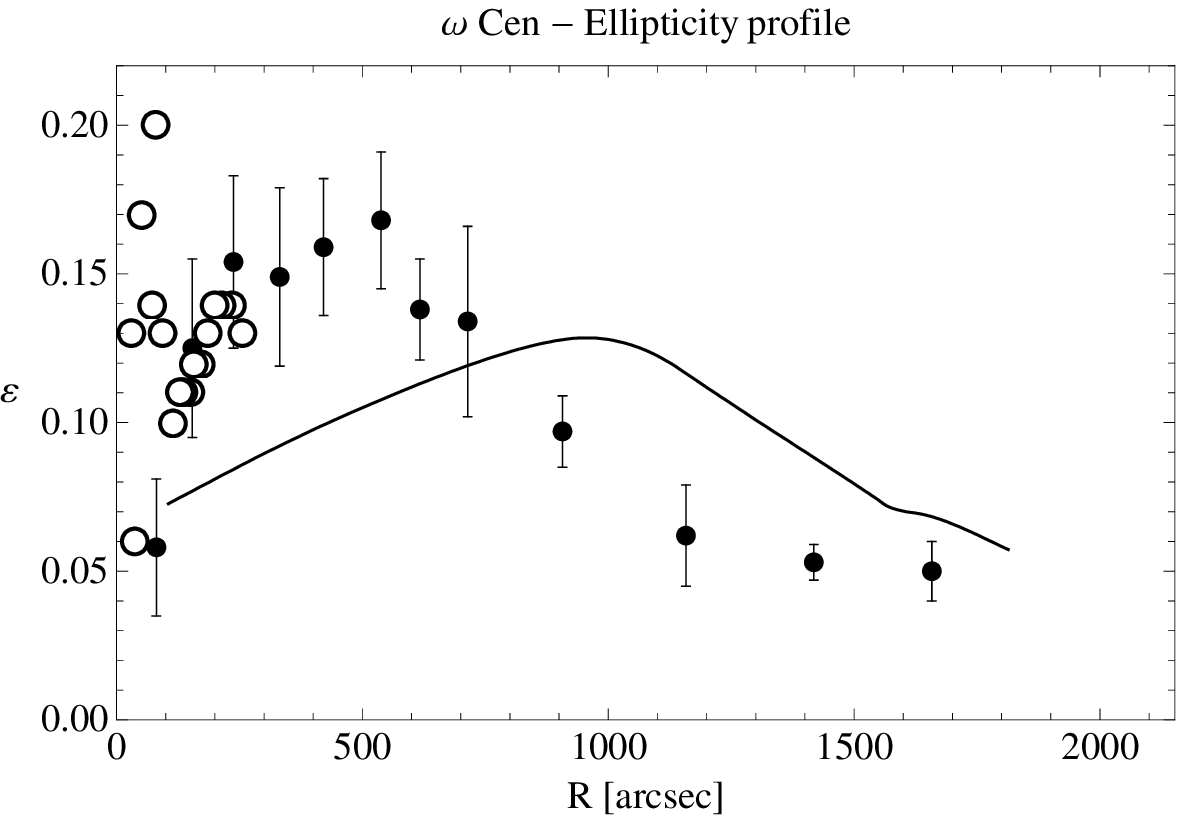}
\caption{\footnotesize
Line-of-sight rotation profile measured along the major axis (top panel) and ellipticity profile (bottom panel) for $\omega$ Cen. Solid lines represent the model profiles, the open circles the observed ellipticities from \citet{Anderson2010}, the black dots those from \citet{Geyer1983}.
}
\label{omega}
\end{figure}

The kinematic data sample for 47 Tuc consists of 2476 line-of-sight-velocities covering the entire extent of the cluster from \citet{Gebhardt1995} and \citet{Lane2011} and 12\,974 HST proper motions from \citet{McLaughlin2006} limited to the central 4 core radii.

The photometric and kinematic profiles are well reproduced by the model, both along the line of sight and on the plane of the sky, confirming isotropy in velocity space in the central region. In particular, the rotation profile is well matched throughout the radial extent of the cluster and the observed ellipticity profile follows the prediction of the model with surprising accuracy (Fig. \ref{47tuc}). Given the fact that the selected model is associated with an ellipticity profile that is the morphological counterpart to the presence of rotation, we conclude that the observed deviations from sphericity originate from the presence of internal rotation.

Finally, we estimate a dynamical distance to 47 Tuc of d=$4.15\pm0.07$ kpc, consistent with previous dynamical estimates, and slightly smaller than the distance derived from photometric methods \citep{McLaughlin2006}.

\section{$\omega$ Cen: a partially relaxed globular cluster}

\begin{figure*}[t!]
\centering
\includegraphics[width=0.65\textwidth]{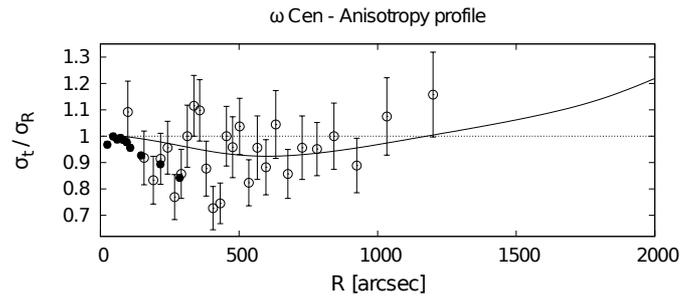}
\caption{\footnotesize
Anisotropy profile of $\omega$ Cen, defined as the ratio of the tangential to the radial component of the proper motion dispersion profile. The solid line represents the model profile, open circles the observational data from \citet{vanLeeuwen2000}, and the black dots the data from \citet{Anderson2010}.}
\label{ani_omega}
\end{figure*}

The kinematic data sample for $\omega$ Cen consists of 1868 line-of-sight-velocities from \citet{Reijns2006} and \citet{Pancino2007}, reaching a radial extent of approximately half truncation radius, 2740 ground-based proper motions from \citet{vanLeeuwen2000}, and 72\,970 HST proper motion measurements from \citet{Anderson2010}, with radial extent of half truncation radius.

The surface brightness and the line-of-sight kinematic profiles are in satisfactory agreement with the model and the predicted average ellipticity is consistent with the one observed. However, discrepancies are noted for the ellipticity profile (Fig. \ref{omega}). We argue that the significant offset between observed and predicted ellipticity profile confirms the complex nature of $\omega$~Cen, resulting from its condition of partial relaxation, and brings out the interplay between anisotropy and rotation. In fact, by analyzing the anisotropy profile derived from the observed proper motions (Fig. \ref{ani_omega}), we note a degree of radial anisotropy higher than predicted. In fact, it had already been noted that radially-biased anisotropic models appear to perform better in this cluster (for example, see the application of the f$^{(\nu)}$ models by \citealp{Zocchi2012} and rotating \citealp{Wilson1975} models by \citealp{Sollima2009}). Yet, our model incorporates the feature of tangential anisotropy observed in the outer parts (considered to be a natural result of the dynamical evolution of a stellar system within an external tidal field) and previously exhibited only by orbit-based dynamical studies \citep{vandeVen2006}.


\section{Conclusions}

We have presented a self-consistent global interpretation of the structure and dynamics of $\omega$ Cen and 47 Tuc, with particular attention to internal rotation, relaxation conditions, and deviations from spherical symmetry. High priority has been given to the interpretation of the available three-dimensional kinematic data.

Internal rotation has been shown to play an important role in determining the structure of the clusters. The different degrees of success of our models have been argued to correspond to the different relaxation conditions of these systems: the discrepancies noted for $\omega$ Cen are explained by the interplay between rotation and a high degree of radial anisotropy.

The use of proper motions in a dynamical analysis has been found to be of primary importance. By matching the observed proper motion profiles with the profiles predicted by our models, we have obtained dynamical estimates of the distances to the two clusters.



\bibliographystyle{aa}

\end{document}